\documentclass[final]{aipproc}
\layoutstyle{6x9}
\begin{document}
\title{Relativistic free-particle quantization on the light-front:
New aspects}
\author{J.H.O.Sales}{
  address={Centro de Ci\^{e}ncias-Universidade Federal de Itajuba, 37500-000 - MG - Brazil}
   ,altaddress={henrique@fatecsp.br}
}
\author{A.T. Suzuki}{
  address={Department of Physics, North Carolina State
University, Raleigh, NC 27695-8202}
}
\author{G.E.R. Zambrano}{
  address={Instituto de F\'{\i}sica Te\'{o}rica - UNESP
Rua Pamplona 145, 01405-900 - SP - Brazil}
}

\begin{abstract}
We use the light-front machinery to study the behavior of a relativistic
free particle and obtain the quantum commutation relations from the
classical Poisson brackets. We argue that the usual projection onto the
light-front coordinates for these from the covariant commutation ralations
does not reproduce the expected results.  
\end{abstract}
\maketitle

\section{Light-front quantization}
The Lagrangian for a free relativistic particle of mass $m$, is given by 
\cite{2} ${\cal L}=-m\sqrt{\stackrel{\cdot }{x}^{2}}$. The canonical four
momentum $p^{\mu }$ is obtained directly from 
\begin{equation}
p^{\mu }=\frac{\partial {\cal L}}{\partial \stackrel{\cdot }{x_{\mu }}}=%
\frac{-m\stackrel{\cdot }{x}^{\mu }}{\sqrt{\stackrel{\cdot }{x}^{2}}}
\label{1.2}
\end{equation}
and the fundamental Poisson brackets are: 
\begin{equation}
\left\{ x^{\mu },p_{\nu }\right\} =\delta _{\nu }^{\mu }
\end{equation}
and
\begin{equation}
\left\{ x^{\mu },x_{\nu }\right\} =0=\left\{ p^{\mu },p_{\nu }\right\} 
\label{1.3}
\end{equation}
Canonical quantization can not be obtained directly from these brackets
through the rules of quantization, because this theory has constraints \cite
{3}. From (\ref{1.2}), we get immediately that 
\begin{equation}
\phi _{1}=p^{2}-m^{2}\approx 0  \label{1.4}
\end{equation}
Using this basic result in 
\begin{equation}
\left\{ x^{\mu },p^{2}-m^{2}\right\} =\left\{ x^{\mu },p^{2}\right\}
-\left\{ x^{\mu },m^{2}\right\}  \label{2.3}
\end{equation}
we have (observing that the second term on the right hand side yields zero
straightforwardly) 
\[
\left\{ x^{\mu },p^{2}\right\} =2p^{\nu }\left\{ x^{\mu },p_{\nu }\right\} , 
\]
which with (\ref{1.3}) gives, 
\begin{equation}
\left\{ x^{\mu },p^{2}\right\} =2p^{\mu }.  \label{2.4}
\end{equation}
For the component $\mu =0$ we have 
\begin{equation}
\left\{ x^{0},p^{2}\right\} =2p^{0}=2E,  \label{2.4a}
\end{equation}
that is, for this particular time component the Poisson bracket relates to
the total energy of the system. In the case of $\mu =+$ coordinate we have 
\begin{equation}
\left\{ x^{\mu },p^{2}\right\} ^{lf}=p^{+}g^{+-}=2p^{+}  \label{2.9a}
\end{equation}
This result agrees perfectly with the covariant case (\ref{2.4}). 
From Eq.(\ref{1.4}) transformed into the light front 
\begin{equation}
\phi _{1}=p^{+}p^{-}-p^{\perp 2}-m^{2}\approx 0  \label{2.11}
\end{equation}
we have the primary constraint.
The light front Lagrangian is 
\begin{equation}
{\cal L}_{lf}=-m\sqrt{\stackrel{\cdot }{x}^{\mu }\stackrel{\cdot }{x}%
_{\mu }}=-m\sqrt{\stackrel{\cdot }{x}^{+}\stackrel{\cdot }{x}^{-}-%
\stackrel{\cdot }{x}^{\perp 2}}\,.  \label{2.12}
\end{equation}
The canonically conjugate momentum components are 
\begin{equation}
p^{-}=-m\frac{\stackrel{\cdot }{x}^{-}}{\sqrt{\stackrel{\cdot }{x}^{2}}}%
\;, p^{+}=-m\frac{\stackrel{\cdot }{x}^{+}}{\sqrt{\stackrel{\cdot }{x}%
^{2}}}
\end{equation}
and 
\begin{equation}
p^{\perp }=-m\frac{\stackrel{\cdot }{x}^{\perp }}{\sqrt{%
\stackrel{\cdot }{x}^{2}}}.
\end{equation}
The Hamiltonian is 
\begin{equation}
H_{c}^{lf}=p\stackrel{\cdot }{x}-{\cal L}_{lf}=0
\end{equation}
This result gives us an indication that we use 
\[
\widetilde{H}^{lf}=\lambda \left( p^{+}p^{-}-p^{\perp 2}-m^{2}\right)
\]
which by the condition that the constraint does not evolve in time we have
\begin{equation}
\stackrel{\cdot }{\phi}_{1}=\left\{ {\phi}_{1},\widetilde{H}^{lf} \right\}
\approx 0 . 
\end{equation}
This means that there are no new constraints and we are unable to determine
the multiplier $\lambda $.

Let us then, as before impose a new constraint in the light front 
\begin{equation}
\phi _{2}=x^{+}-s\approx 0 \,,  \label{2.15}
\end{equation}
so that, 
\begin{equation}
\left\{ \phi _{1},\phi _{2}\right\} =-2p^{+}  \label{2.16}
\end{equation}

Comparing with the result (\ref{2.4a}) we again perceive that there is an
inconsistency here. Instead of the energy $p^{-}$ we get the momentum $p^{+}$%
.

Constructing the Hamiltonian
and with the condition of non evolution in time of the constraint
we have 
\begin{equation}
H=\frac{1}{2p^{+}}\left( p^{+}p^{-}-p^{\perp 2}-m^{2}\right)   \label{2.17}
\end{equation}

In order to obtain the Dirac brackets, let us construc the matrix $C$, where 
$C_{ij}=\left\{ \phi _{i},\phi _{j}\right\} $ and $i,j=1,2$. The Dirac
brackets then is given by 
\[
\left\{ x^{\mu },p^{\nu }\right\} _{D}^{lf}=g^{\mu \nu
}-\left\{ x^{\mu },\phi _{1}\right\} C_{12}^{-1}\left\{ \phi _{2},p^{\nu
}\right\} -\left\{ x^{\mu },\phi _{2}\right\} C_{21}^{-1}\left\{ \phi
_{1},p^{\nu }\right\} . 
\]

Quantization is now achieved by 
\begin{equation}
\left[ x^{\mu },p^{\nu }\right] _{D}^{lf}=i\left[ g^{\mu \nu
}-\left( p^{+}g^{\mu -}+g^{\mu +}p^{-}-2p^{\perp }g^{\mu \perp }\right) 
\frac{1}{2p^{+}}g^{+\nu }\right]  \label{2.22}
\end{equation}
Clearly we have here the problem of zero modes in the light-front.
\section{Conclusion}
We have (\ref{2.4}) for the component $\mu =0$ we have $\left\{
x^{0},p^{2}\right\} =2p^{0}=2E,$ that is, for this particular time component
the Poisson bracket relates to the total energy of the system. For the
light-front case, $\mu =+$ and therefore $\left\{ x^{+},p^{2}\right\}
=2p^{+}.$ Which means that the light-front ``time'' variable $x^{+}$ relates
to the momentum $p^{+}$, an apparent inconsistency between canonically
conjugate variables. In the covariante case we have a correlation between
time and energy in the Poisson brackets, while in the light front this
correlation is lost. See Eq.(\ref{2.9a}). This conclusion can be the origin
to problem of zero modes in the light-front. In the paper \cite{5}, we propose a treatment for the presented inconsistency

\begin{theacknowledgments}
J.H.O.Sales thanks the hospitality of Instituto de F\'{i}sica Te\'{o}rica/UNESP. 
A.T.Suzuki acknowledges partial support from CNPq (Bras\'{i}lia)and CAPES (Bras\'{i}lia) and G.E.R.Zambrano acknowledges total support from CAPES
(Bras\'{i}lia).

\end{theacknowledgments}
\bibliographystyle{aipprocl}

\end{document}